\documentclass[a4paper,twocolumn,twoside,10pt]{article}
\newcommand{\beq}{\begin{equation}}
\newcommand{\eeq}{\end{equation}}
\usepackage{graphics}
\usepackage{epsf}
\usepackage{epsfig}
\usepackage{amsmath}
\usepackage{amsfonts}
\usepackage{amssymb}
\usepackage{graphicx}
\usepackage{epsfig}
\usepackage{bm}
\usepackage{authblk}

\begin{document}

\title{Giant texturing effect in multiferroic MnWO$_4$ polycrystals}

\author{Somnath Jana,$^{1,^\dagger}$ Abhishek Nag,$^{2}$ and Sugata Ray$^{1,2}$}

\affil{{\small\it $^1$Centre for Advanced Materials, Indian Association for the Cultivation of Science, \\ \small\it Jadavpur, Kolkata 700 032, India}}
\affil{{\small\it $^2$Department of Materials Science, Indian Association for the Cultivation of Science, \\ \small\it Jadavpur, Kolkata 700 032, India}}

\maketitle



\begin{abstract}
Different methods of texturing polycrystalline materials are developed over years to use/probe anisotropic material properties with relative ease, where complicated and expensive single crystal growth processes could be avoided. In this paper, particle morphology assisted texturing in multiferroic MnWO$_4$ has been discussed. Detailed powder x-ray diffraction vis-a-vis scanning electron microscopic studies on differently annealed and processed samples have been employed to probe the giant texturing effect in powdered MnWO$_4$. A quantitative measure of the texturing has been carried out by means of Rietveld analysis technique. Qualitative presentation of magnetic  and dielectric data on textured pellet demonstrated the development of clear anisotropic physical properties in polycrystalline pellets. Finally, we established that the highly anisotropic plate like particles are formed due to easy cleavage of the significantly large crystalline grains.
\end{abstract}


\let\thefootnote\relax\footnotetext{Supported by DST Fast Track, India and CSIR, India for fellowship to SJ and AN.}

\let\thefootnote\relax\footnotetext{$^\dagger$ Author to whom correspondence should be addressed. e-mail: camsj@iacs.res.in}

\newpage

\section{Introduction}%
Single crystals by definition are the purest form of solid materials and naturally the most exclusive form for carrying out direction dependent bulk experimental studies or for technological uses. However, entropy related effects often facilitate presence of imperfections in the microstructure of solids, such as impurities, inhomogeneous strain or crystallographic defects, and as a result perfect single crystals of reasonable volume are extremely difficult to produce in the laboratory and even then, it might be possible only under precisely controlled conditions. On the other hand, material `texturing', {\it i.e.} the ability to impose preferential orientation to small, crystalline grains of a polycrystalline material, could provide a rather useful and viable solution to this problem. As most of the crystallographic properties are coupled with the crystallographic directions, attempt of texturing could help in realizing many desired, direction specific properties and also in probing anisotropic properties in an otherwise polycrystalline sample. Obviously, if the polycrystalline ceramics with properties close to single crystal can be fabricated by `texturing', they would be advantageous due to their ease of fabrication, shaping, and cost-effectiveness. Following this idea, various experimental techniques have been developed in last few decades to reinforce preferential orientation in materials having versatile anisotropic properties. For example, textured $\alpha$-Al$_2$O$_3$ based ceramics have been made by a number of techniques, {\it e.g.} tape casting,$^{~\cite{Dimarcello}}$ templated grain growth (TGG) {\it etc.},$^{~\cite{Brandon}}$ while Matsuzawa {\it et al.} reported the TGG-type directed growth of ferrites in 1982.$^{~\cite{Matsuzawa}}$ Reactive templated grain growth (RTGG) is another industrial development, which has been utilized to prepare various textured ferroelectric ceramics like Bi$_4$Ti$_3$O$_{12}$,$^{~\cite{Takenaka}}$ CaBi$_4$Ti$_4$O$_{15}$,$^{~\cite{Takeuchi}}$ or pyroelectrics ({\it e.g.} (ZnO)$_5$In$_2$O$_3$,$^{~\cite{Tani}}$ NaCo$_2$O$_4$,$^{~\cite{Tajima}}$) as well as piezoelectric ceramics ({\it e.g.} K$_{0.5}$Na$_{0.5}$NbO$_3$).$^{~\cite{Zang}}$ Texturing of magnetic materials is also demonstrated at high temperature by solidification under an applied magnetic field.$^{~\cite{Rango}}$

Interestingly, certain materials may naturally promote `texturing' in powder forms, which can on the other hand create difficulties because absence of prior information about the degree of such natural `texturing' may mislead researchers while describing physical properties of polycrystals of such materials. But, a more general and questionable issue is the extraction of crystallographic information of such materials through powder diffraction experiments, a method employed to all samples grown either as single crystal or polycrystal, because the diffraction data are normally analyzed with an inherent assumption that the experimental results are average, isotropic responses of the material, which obviously turns invalid if the powder becomes strongly textured. In this paper, we report that substantial natural `texturing' occurs in the well-known multiferroic MnWO$_4$$^{~\cite{Taniguchi,Heyer,Arkenbout}}$ in its powder form. Therefore, all the powder diffraction data providing crystallographic information, inherently connected to its multiferroic property, should be analyzed very carefully with detailed, prior knowledge about the `texturing' that may remain in the sample under study. Moreover, a fair body of research on polycrystalline MnWO$_4$ also exist$^{~\cite{ref2}}$ which naturally involve serious uncertainties. On the other hand, this natural texturing tendency of the grains in polycrystalline MnWO$_4$ could be consciously exploited to create nearly single crystalline properties and single crystal-like anisotropies in MnWO$_4$ polycrystals. Also, the high sensitivity of the electrical conductivity of MnWO$_4$ to the level of surrounding humidity makes it a potential system for designing humidity sensors,$^{~\cite{Dellwo, Wenmin}}$ and the tendency of texturing could be exploited for making single crystal-like films easily, which could be useful for application.$^{~\cite{Wenmin1}}$

\section{Experimental Methods}
Polycrystalline MnWO$_4$ was synthesized using solid state reaction method. Stoichiometric amounts of MnO (99.9\%) and WO$_3$ (99.9\%) were mixed in an agate mortar and prolonged grinding was carried out in C$_2$H$_5$OH medium to ensure maximum possible chemical homogeneity in the starting mixture. The resultant powder was then thermally treated in different conditions to study the grain growth and texturing; the details are shown in {\bf Fig. 1}. Also, MnWO$_4$ single crystals were grown by the floating zone technique. The phase purity was checked by powder x-ray diffraction method in a Bruker AXS: D8 Advance x-ray diffractometer and the sample morphologies were probed by scanning electron microscopy (SEM) in a JEOL JSM-6700F FESEM instrument. The magnetic measurements were done in a Quantum Design SQUID magnetometer, while an indigenously built laboratory setup with TEGAM 3550 LCR meter was used for dielectric measurements.

\begin{figure}
\resizebox{7cm}{!}
{\includegraphics*[126pt,382pt][465pt,690pt]{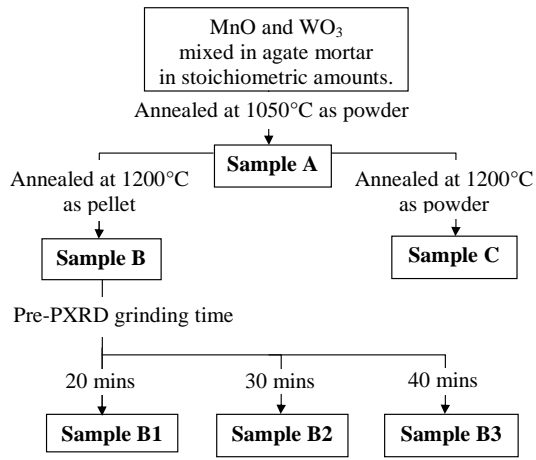}} \\
\caption{Hierarchy chart showing annealing conditions for MnWO$_4$ samples A, B and C and pre-PXRD grinding times for samples B1, B2 and B3.}
\label{fgr:Fig1}
\end{figure}

\section{Results and discussions}%
To have a quantitative idea of the anisotropy in grain shapes, we simply applied manual pressure on the powder during the sample arrangement for powder x-ray diffraction (PXRD) measurements, thus enforcing preferential orientation of anisotropically shaped grains,$^{~\cite{P}}$ if present. All the peaks of PXRD patterns from all the samples could be indexed perfectly with the $P2/c$ space group with the crystallographic parameters $a$ = 4.829 \AA, $b$ = 5.758 \AA, $c$ = 4.996 \AA, and $\beta$ = 91.146$^{\circ}$, which confirm phase purity of all the samples. In {\bf Fig. 2(a)}, the PXRD pattern from sample A (open circles) is shown along with the standard powder data (red columns) from literature. Before collecting this data, a chunk of the fine powder was pressed and flattened on the sample holder by a glass slide. Although the PXRD pattern matches with the literature data at a first glance, a closer look reveals that the relative intensities of a few crystalline peaks differ rather strongly from the standard data with the strongest discrepancies corresponding to the planes parallel to (0 1 0) Miller index (see the insets to Fig. 2(a)), indicating a preferential orientation of grains along this direction. The SEM image of the same pressed sample is also shown in the inset. The particles are nearly spherical with few open facets, which are expected to be along the (0 1 0) planes.
\begin{figure}[h]
\resizebox{7cm}{!}
{\includegraphics*[92pt,420pt][285pt,691pt]{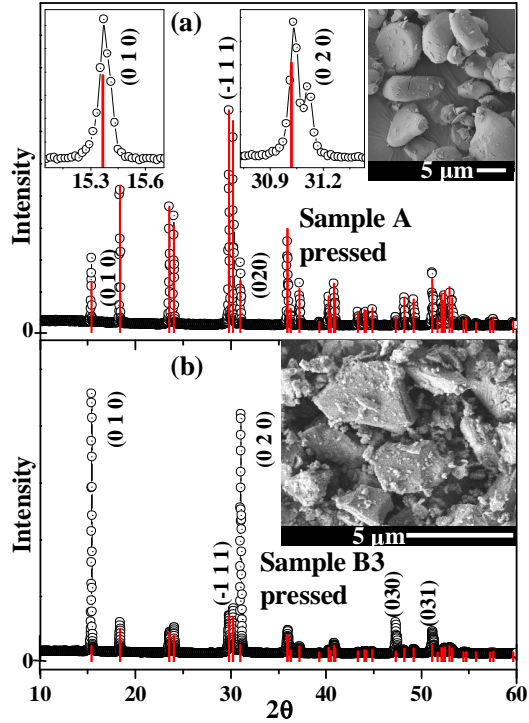}} \\
\caption{(Colour Online) X-ray powder diffraction from pressed MnWO$_4$ samples: A(panel(a))and B3(panel(b)) at 300 K displaying the observed data (open circles) and standard data from literature (red columns), along with insets showing corresponding SEM images and enlarged (0 1 0) and (0 2 0) reflections.}
\label{fgr:Fig2}
\end{figure}
Usually almost all powder samples, due to the presence of certain level of anisotropy in their grain shapes are more or less affected by preferred orientation,$^{~\cite{Wenk}}$ but the gigantic effect observed in high temperature annealed MnWO$_4$ samples (Fig. 2(b)) is unprecedented. In this case, the sample was annealed at 1200~$^{\circ}$C in the form of a pellet (sample B), thoroughly ground to fine powder (equivalent to sample B3, Fig. 1) and then placed onto a diffraction sample holder under manual pressure. Evidently, giant enhancements in the peak intensities could be observed at reflections (0 1 0), (0 2 0), (0 3 0), (0 3 1) {\it etc.} which has to be extrinsic and should be an outcome of heavy texturing. Consistent with this assumption, the SEM experiments (see inset,panel (b)) exhibit plate-like grains for sample B3.

\begin{figure}[h]
\resizebox{7cm}{!}
{\includegraphics*[70pt,51pt][536pt,762pt]{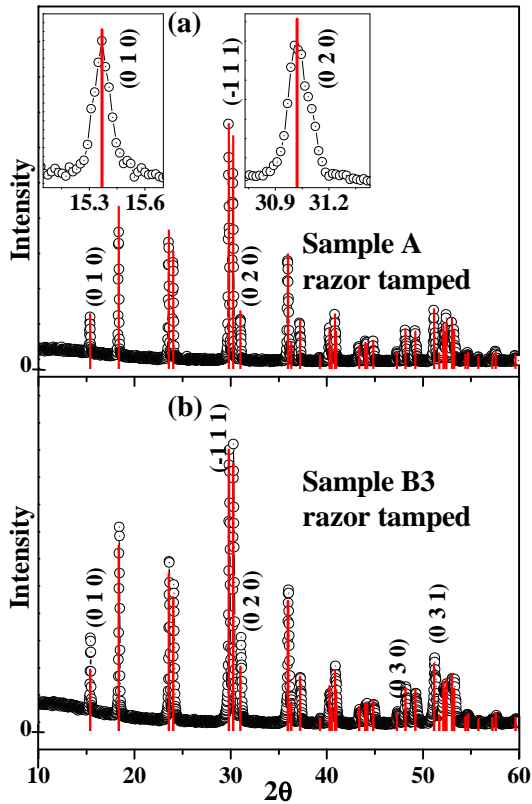}} \\
\caption{(Colour Online) X-ray powder diffraction from `razor tamped' MnWO$_4$ samples: A(panel(a))and B3(panel(b)) at 300 K showing the observed data (open circles), standard data from literature (red columns) and enlarged (0 1 0) and (0 2 0) reflections as insets.}
\label{fgr:Fig3}
\end{figure}

In order to check the true PXRD patterns of these samples, we employed the simplest known method to avoid preferred orientation in sample preparation for PXRD, namely the {\bf `razor tamped surface' (RTS)} method and obtained random particles at the surface of the measurement holder.$^{~\cite{razor}}$ Here, the sample surface was tamped by the sharp edge of a razor blade during sample preparation, and the PXRD results for samples A and B3 under this new preparation have been plotted in {\bf Fig. 3}. It is evident from Fig. 3 that PXRD patterns for both the samples showing correct relative intensities could be reverted using this method. Therefore, it is established that the grains of sample B3 are highly anisotropic and can be textured very easily and the lack of prior knowledge about the morphology of polycrystalline MnWO$_4$ being studied, may lead to misleading diffraction or other experimental data.$^{~\cite{thesis}}$

As mentioned earlier, this high texturing tendency of polycrystalline MnWO$_4$ could on the other hand be exploited to an important advantage as well. Careful processing of these polycrystals could generate nearly single crystalline anisotropy and related properties, which should be of significant help to material physicists or technologists. But to make use of the texturing one first needs to quantify the extent of grain alignments in MnWO$_4$ which we did next by applying small manual pressures to the samples and carefully analyzing the experimental PXRD patterns. Now, in order to have such quantitative idea about the particle shape and the fraction of textured particles, the PXRD pattern of the compressed sample B3 was fitted with the help of modified March's function as implemented in the {\it Fullprof} program.$^{~\cite{manual, Dollase}}$ The experimentally observed, calculated and difference profiles are presented in {\bf Fig. 4} with (panel (b)) and without (panel (a)) taking into account the preferential orientation along (0 1 0) while fitting. It is to be noted that the value of R$_{wp}$ is reduced to 17.9\% from 72.6\% after adding the preferred orientation along (0 1 0) along with a large improvement in fitting, indicating the presence of high preferential orientation of the grains along the (0 1 0) planes. The result of the fitting indicates highly plate-like morphology of the grains and as high as $\sim$50\% grain texturing along the (0 1 0) planes.
\begin{figure}[h]
\resizebox{7cm}{!}
{\includegraphics*[73pt,53pt][537pt,761pt]{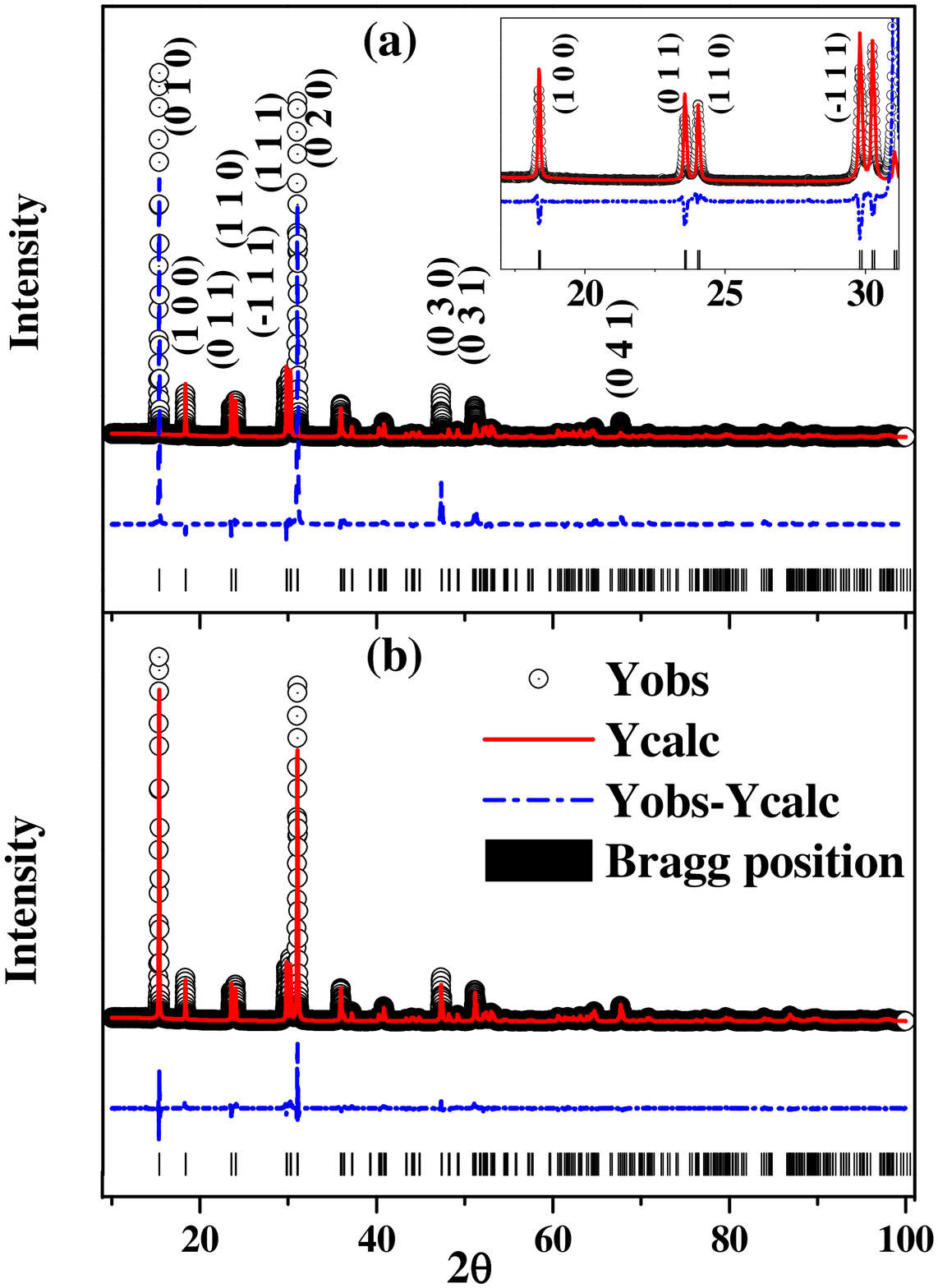}} \\
\caption{(Colour Online) X-ray powder diffraction from pressed MnWO$_4$ sample B3 at 300 K showing the observed data (open circles), calculated pattern ( red continuous line), difference pattern (blue broken line) and Bragg positions with preferred orientation along (0 1 0) considered in (panel(b)) and not in (panel (a)).}
\label{fgr:Fig4}
\end{figure}

\begin{figure}[h]
\resizebox{7cm}{!}
{\includegraphics*[85pt,100pt][552pt,718pt]{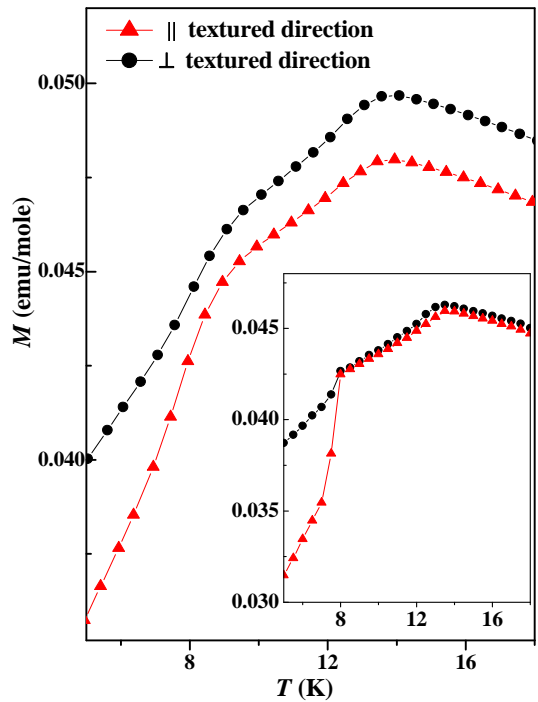}} \\
\caption{(Colour online) Magnetization $M$($T$) data observed (main panel) and simulated (inset) for MnWO$_4$ pellet (equivalent to sample B3) with the magnetic field parallel (black circle) and perpendicular (red triangle) to the flat pellet surface.}
\label{fgr:Fig5}
\end{figure}
This amount of texturing would then surely be manifested through the anisotropic physical properties of a sample. Using magnetic and dielectric measurements, here we illustrate the apparent anisotropy in the properties of polycrystalline MnWO$_4$ pellets. These pellets are made employing unidirectional pressure in a pressure die, which enforce substantial texturing in the pellets where at least 50\% of the particles (equivalent to sample B3) are expected to be textured having the crystallographic $b$ direction perpendicular to the flat pellet surface. The magnetic moments with varying temperatures were measured on such a sintered polycrystalline pellet mounted in two different configurations with respect to the magnetic field direction. To compare the measured anisotropy, a $M$($T$) data was simulated by a linear superposition of the same from a reported single crystal$^{~\cite{Taniguchi}}$ and the measured $M$($T$) data from bare powders, so that an overall 50\% texturing is ensured. In {\bf Fig. 5} the measured (main panel) and simulated (inset) $M$($T$) data are plotted for the two aforementioned directions. The clear similarity between the two offers the evidence for the presence of large anisotropy in the polycrystalline pellet due to large texturing effects.

\begin{figure}[h]
\resizebox{7cm}{!}
{\includegraphics*[41pt,109pt][569pt,757pt]{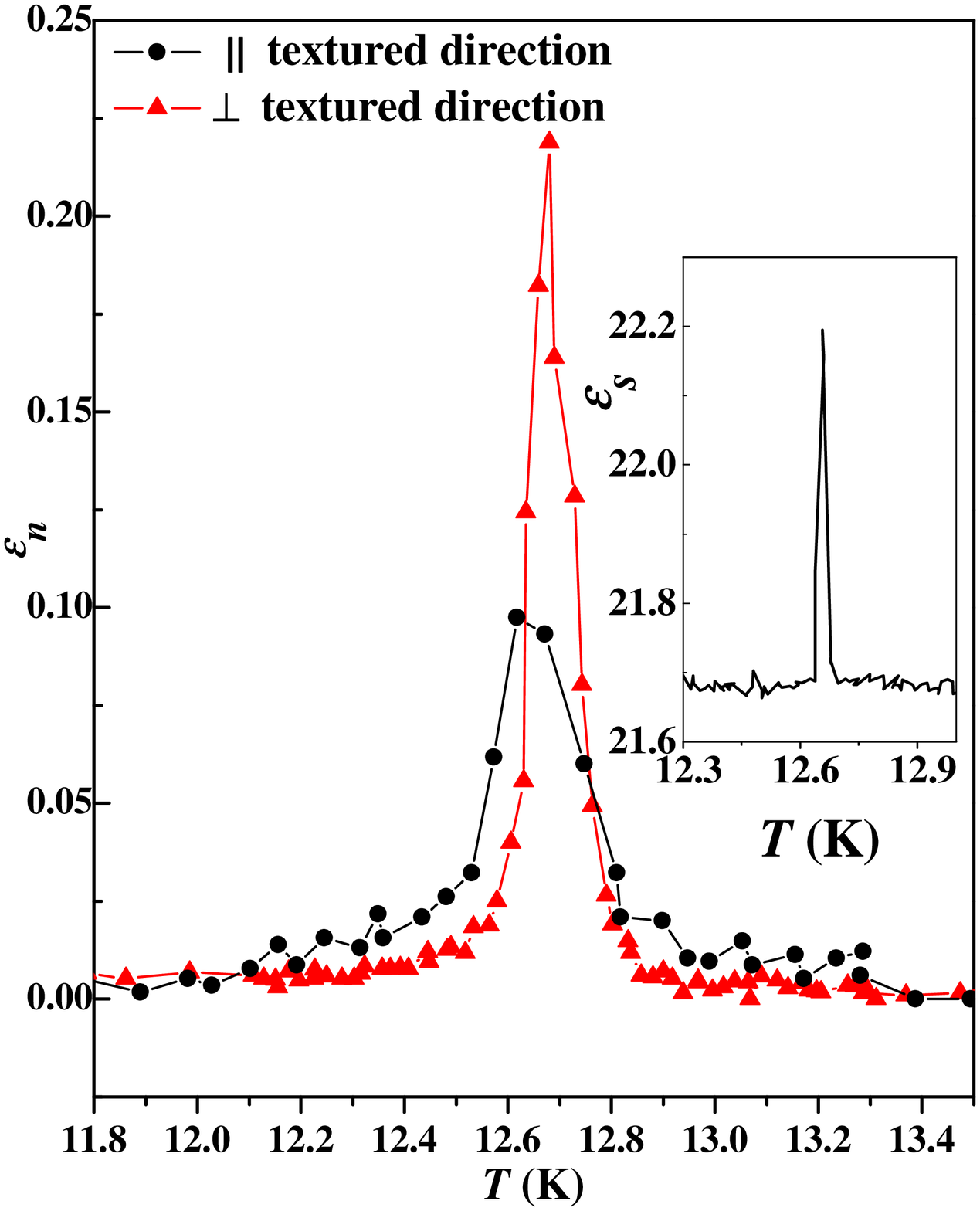}} \\
\caption{(Colour online) Normalized relative dielectric constants measured with changing temperature for MnWO$_4$ pellet (equivalent to sample B3) along (black circle) and perpendicular (red triangle) to the flat pellet surface. Inset shows the value obtained for a single crystal of MnWO$_4$.}
\label{fgr:Fig6}
\end{figure}

The variation in dielectric constants with temperature were measured on a similarly made pellet, once across the flat surface and once perpendicular to it. Dielectric constant of a single crystal of MnWO$_4$ was also measured in order to make a comparison with the values obtained from the pellet. {\bf Fig. 6} shows the normalized values of relative dielectric constants for the two directions of the polycrystalline pellet and the data obtained for the single crystal in the inset. The normalized values ($\epsilon^n$) have been calculated using
$$\epsilon^n=\frac{\epsilon_m-\epsilon_b}{\Delta\epsilon_s}$$
where, $\epsilon_m$, $\epsilon_b$ and $\Delta\epsilon_s$ are the measured dielectric constants of the polycrystalline pellet, the corresponding background values and the peak height of the single crystal respectively. Expectedly, the peak heights from the polycrystalline pellet are substantially lower compared to the single crystal (maximum $\epsilon^n$ being 0.25) but more importantly, the clear contrast in the $\epsilon^n$ peak heights for the two different directions brings out yet another proof of single crystal like anisotropy generated in a polycrystalline sample by texturing.
\begin{figure}[h]
\resizebox{7cm}{!}
{\includegraphics*[14pt,41pt][580pt,782pt]{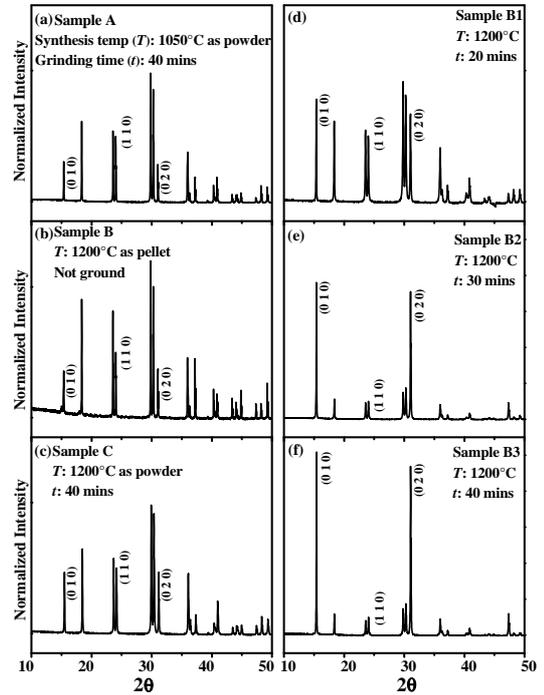}} \\
\caption{X-ray powder diffraction ($I_{(1 1 0)}$ = 1.0 for all the panels) from pressed MnWO$_4$ samples A, B, C, B1, B2 and B3 in panels (a)-(f) respectively. Panels (e) and (f) have been scaled up to accommodate the large preferred orientation effect.}
\label{fgr:Fig7}
\end{figure}

Next, it is important to understand the formation mechanism of the preferentially oriented plate-like grains in polycrystalline MnWO$_4$ for further uses. We have carried out a few controlled measurements to figure out the aligned grain formation mechanism. In {\bf Fig. 7}, x-ray diffraction patterns from samples A (annealed at 1050~$^{\circ}$C), B (annealed at 1200~$^{\circ}$C in pellet form), C (annealed at 1200~$^{\circ}$C in powder form), B1, B2 and B3 are shown in different panels. The pre-PXRD sample preparations however, were different for different samples. Samples A and C, annealed as powders, were thoroughly ground before collecting the PXRD, while in case of sample B, the hard annealed pellet was just broken into powder and any severe grinding was avoided. On the other hand, samples B1, B2 and B3 were prepared from sample B with increasing grinding times (see Fig. 1). All the samples were then placed on the sample holder and pressed with glass slides before measurement, so that, should aligned grains be already present in any of them, the texturing effect would show up in PXRD. Interestingly, the XRD patterns reveal that preferential alignment is absent in samples A, B, and C, refuting the idea of any preferential grain growth in these samples.
\begin{figure}[h]
\resizebox{7cm}{!}
{\includegraphics*[83pt,446pt][437pt,730pt]{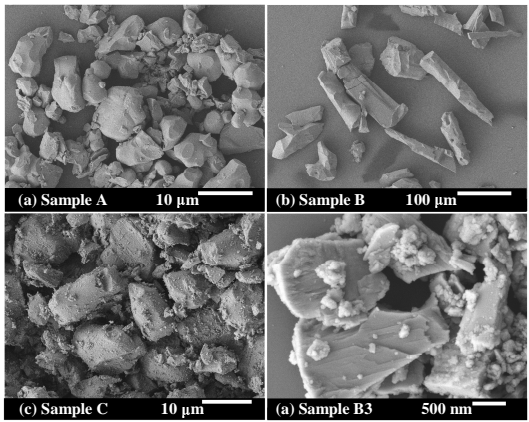}} \\
\caption{Scanning electron microscopy images from pressed MnWO$_4$ samples A, B, C and B3 in panels (a)-(d) respectively showing sample morphology.}
\label{fgr:Fig8}
\end{figure}

One must mention at this point that a proper comparison of the degree of texturing in these samples is an exceedingly hard process because nearly all the PXRD peaks get affected under the influence of significant texturing, making the pattern normalization at any certain invariant 2$\mathrm \theta$ peak nontrivial. However, a careful look into Fig. 4(a) tells that the intensities of the observed and calculated profiles match better at (1 1 0) compared to (1 0 0), (0 1 1), (-1 1 1), (1 1 1) (see inset to {\bf Fig. 4(a)}), among the reflections those are not parallel to [0 1 0]. From the consideration of the preferred orientation function,$^{~\cite{Book}}$ it is understood that the (1 1 0) group of planes will indeed be the least affected ones, while there would be significant intensity variations for all other groups of planes. Thus in order to compare different levels of preferred orientation, the patterns were normalized at (1 1 0), where the effect is minimal. From {\bf Fig. 7(c), (d) and (e)}, one can clearly see that the intensities at (0 1 0), (0 2 0) and other related reflections are increasing with increasing grinding times. The PXRD patterns in Fig. 7 thus indicate that the as-formed sample B does not exhibit texturing and all the anisotropic crystallites originate only as a result of mechanical grinding.

SEM images collected from samples A, B, C and B3 and are shown in {\bf Fig. 8}. The SEM image from sample B stands out from A and C, as huge single crystal like, faceted grains are seen. We have already discussed earlier that it is only sample B which exhibits significant texturing effects after thorough grinding indicating that the formation of large crystallite grains is a pre-requisite for texturing. The SEM image from sample B3 {\it i.e.} the portion of sample B which underwent maximum grinding shows large flat plates. Therefore, one could conclude that large crystalline grains of MnWO$_4$ (sample B) are cleaved along preferential faces during mechanical grinding (samples B1, B2, and B3), which at the end gives rise to strong texturing effects.

\section{Conclusions}
MnWO$_4$ exhibits a natural cleaving tendency and hence texturing. Careful treatment of polycrystalline MnWO$_4$ can produce aligned ingots where the crystalline anisotropy could be as high as 50\%. Our results show that large crystallites, ground heavily to form powders, create plate-like highly anisotropic grains, which can give rise to huge texturing under application of small manual pressures. Hence, powder diffraction experiments carried out on crushed single crystals or highly dense polycrystalline ingots of MnWO$_4$ could be highly prone to such effects and as a result can provide largely misleading crystallographic information. The magnetic and dielectric properties measured along two perpendicular directions on a pressed pellet show clear development of the anisotropic materials properties, which further ensure the morphology induced texturing in MnWO$_4$.

\end{document}